# Detecting causality in Plant electrical signal by a hybrid causal analysis approach

Yang Chen [1,3], Dong-Jie Zhao [4], Chao Song [1,3], Wei-He Liu [1,3], Zi-Yang Wang [1,3], Zhong-Yi Wang [1,2,3], Guiliang Tang [4] and Lan Huang [1,3,*]

1. College of Information and Electrical Engineering, China Agricultural University, Beijing 100083, China; hlan@cau.edu.cn
2. Modern Precision Agriculture System Integration Research Key Laboratory of Ministry of Education, Beijing 100083, China;
3. Key Laboratory of Agricultural information acquisition technology (Beijing), Ministry of Agriculture, Beijing 100083, China;
4. Michigan Technological University, Houghton, MI 49931-1295 USA; gtang1@mtu.edu

yancychy@cau.edu.cn, dongjiez@mtu.edu, songchaodevip@cau.edu.cn,

lwh93@cau.edu.cn, s13111175@cau.edu.cn, wzyhl@cau.edu.cn, gtang1@mtu.edu, hlan@cau.edu.cn

* Correspondence: hlan@cau.edu.cn; Tel.: +86-010-6273-7778

**Abstract:** At present, multi-electrode array (MEA) approach and optical recording allow us to acquire plant electrical activity with higher spatio-temporal resolution. To understand the dynamic information flow of the electrical signaling system and estimate the effective connectivity, we proposed a solution to combine the two casualty analysis approaches, i.e. Granger causality and transfer entropy, which they complement each other to measure dynamics effective connectivity of the complex system. Our findings in three qualitatively different levels of plant bioelectrical activities revealed direction of information flow and dynamic complex causal connectives by using the two causal analysis approaches, especially indicated that the direction of information flow is not only along the longitudinal section but also spreading in transection.

Highlight

1. For plant electrical signaling complex system , we proposed a solution to combine the two casualty analysis approaches, i.e. conditional Granger causality and transfer entropy, which they complement each other to measure dynamics effective connectivity of the complex system.
2. Our findings indicated how dynamic effective connectivity in the regions generated when plant bioelectrical activity occurred.

## 1. Introduction

Plant tissues can be seen as a multi-layered complex network, which respond to external stimuli and generate a cascade of physiological and biological reactions. Plant electrical signal is an initial physiological response to the stimuli of environment, and have an important effect on plant physiological activities such as photosynthesis, respiration and growth regulation [1-9]. By considering the electrical excitation of plants, electrical signals in plants can be divided into three types: action potential (AP) and variation potential (VP), system electrical potential (SP), and all the signals can transmit from the stimulated site to other parts of the plant [7-9].

Despite physiological significance of electrical signals in plants, extracting information from signal is challenging due to limitation of conventional recording techniques. Nowadays, multielectrode array (MEA) approach and optical recording allow us to acquire plant electrical activity with higher spatio-temporal resolution [10-13].

Moreover，for plant electrical signals induced by external stimuli, where bioelectrical activities initially occur, which regions are important for spreading? How can causal pathways and effective connectivity in these regions be identified? When the coupling strength of global casual network connectivity significantly enhanced? These questions are important for understanding the dynamics and functioning of the electrical signaling system, even predicting their behavior. Therefore, to solve these problems, causal analysis based on process data is able to find causal dependencies among observed data and thereby obtain the propagation path of disturbances.

Detecting causality between observed time series from multivariate complex system has challenged researchers to determine the couplings, direction, and the quantification of the coupling strengths of causal interaction. Among a number of linear regression-based methods, Granger causality analysis has become increasingly well-known one over recent years. The concept of Granger causality was proposed by Granger in 1969 [14]. In fact, GC is a statistical notion of causal influence based on prediction model via vector autoregression. In practice, it is able to measure causality in bivariate time series, and then applied in multivariate complex system, e.g. identification of functional and effective connectivity in brain from fMRI, MEG and EEG data [15-19]. Because there are always possible effects of common drivers and indirect influences among variables using multivariate GC, conditional Granger causality can reveal direct causal effects among the components of a complex system and exclude spurious connectivity [20-23]. Generally, GC is not only used to measure causality for linear multivariate processes but also extend to deal with non-linear relationships by filtering, detrending and differencing the raw data [24-29]. However, it is worth noting that preliminary data treatment may result in serious distortions [30].

Although Granger causality is undoubtedly a widely used method to detect causal relationship between different factors, it remains challenges when the interaction between two signals is nonlinear and cross-talk between signals, even signals not matching precondition of stationary [31]. To overcome the limitations and quantify causal relationship between the intertwined variables of nonlinear systems, transfer entropy (TE) was introduced by Schreiber to estimate effective connectivity [32]. TE is asymmetric and allows differentiation in the direction of information propagation. Unlike GC, TE is a model-independent method based on information theory without requirement for a model of the interaction [31, 33]. Moreover, there are a lot of extended TE method, e.g. TE with multiple delays in time series and multivariable TE [34-35]. Despite used originally in econometrics, it has since found applications in many fields, e.g. electrophysiological systems [36], chemical processes [37] and industry [38], particularly in neuroscience [39-40]. Lizier et al. established a framework that combined multivariate mutual information (MI) and TE to analyze fMRI time series for revealing direction flow of information between brain regions involved in a visuo-motor tracking task [41]. By using TE, neurophysiological evidence of mutual relationships between brain regions might enhance the understanding of cortico-cortical communication during drowsy driving [42]. In addition, an overview has reported that TE was superior in terms of stability and accuracy when applied to time series from linear and non-linear relations with unidirectional and bidirectional coupling [43].

To date, it remain unclear which interaction relationships among these plant electrical signals, linear or nonlinear relation. For this reason, we proposed a solution to combine the two casualty analysis approaches, i.e. conditional Granger causality and transfer entropy, which they complement each other to measure dynamics effective connectivity of the complex system.

## 2. Granger causality and Transfer Entropy

*2.1 Granger causality*

The Granger causality is a fundamental tool for the description of causal interactions of two signals. Granger causality could be used as a method that to estimate whether those two time series exist direct causality or indirect causality. Let X={x(n)} and Y={y(n)} as time series, in order to calculate their Granger causality relationship, auto-regression method will be used to establish a p-order autoregressive model of X and Y:

$$x(n) = \sum_{i=1}^{p} c_{1i} x(n-i) + e_1(n) \quad (1)$$

$$y(n) = \sum_{i=1}^{p} c_{2i} y(n-i) + e_2(n) \quad (2)$$

where $e_1(n)$ and $e_2(n)$ is prediction errors that only depend on past value of the signals. Bivariate p-order AR model could be used to simulate X and Y in order to build causal model of X and Y.

$$x(n) = \sum_{i=1}^{p} c_{3i} x(n-i) + \sum_{i=1}^{p} c_{4i} y(n-i) + w_1(n) \quad (3)$$

$$y(n) = \sum_{i=1}^{p} c_{5i} y(n-i) + \sum_{i=1}^{p} c_{6i} x(n-i) + w_2(n) \quad (4)$$

Where $w_1(n)$ and $w_2(n)$ is prediction error. They depend on the past value of X and Y.

Let $\sum = \begin{pmatrix} \sum_2 & \Upsilon_2 \\ \Upsilon_2 & \Gamma_2 \end{pmatrix}$, where $\sum_2 = \text{var}(w_1(n))$, $\Gamma_2 = \text{cov}(w_1(n), w_2(n))$.

If X and Y are independent, then coefficients $c_{4i}$ and $c_{6i}$ are zero, and $\Upsilon_2 = 0$.

All parameters could be obtained by solving Yule-Walker equations. Orders of the model could be determined by Akaike information criterion (AIC). The variance of prediction error in AR model for one-dimensional variable X can represent prediction accuracy by

$$\sum\nolimits_{X|X^-} = \frac{1}{N-p} \sum_{i=1}^{N} (e_1(n))^2 \quad (5)$$

For the prediction accuracy of AR model for two-dimensional variable X and Y, the variance of the prediction error is given by

$$\sum\nolimits_{X|X^-Y^-} = \frac{1}{N-2p} \sum_{i=1}^{N} (w_1(n))^2 \quad (6)$$

Accordingly, Granger Causality influence from Y to X is defined by

$$F_{Y \to X} = \ln \frac{\sum_{X|X^-}}{\sum_{X|X^-Y^-}} \quad (7)$$

According to the definition of Granger Causality, if the fitting error in equation (3) after adding Y is smaller than the fitting error in equation (1), then $F_{Y \to X} > 0$, and it indicates that Y cause X. If $F_{Y \to X} = 0$, then there do not exist causality. The causal influence from X to Y can be expressed by

$$F_{X \to Y} = \ln \frac{\sum_{Y|Y^-}}{\sum_{Y|Y^-X^-}} \quad (8)$$

*2.2 Transfer entropy*

The definition of transfer entropy is written in literatures [32]. Given discrete time series X and Y, p(x) and p(y) are their probability distribution functions respectively. Here, p(x, y) represents their joint probability distribution, p(i|j) is their conditional probability.

Assume the state of X at time n is $X_n$, and $X_n^{(k)}$ represents $(X_n, ..., X_{n-k+1})$. If the past of X and Y have an impact on the future of Y, the information flow from X to Y can be expressed by transfer entropy in equation (9).

$$T_{X \to Y}(k,l,u) = \sum p(Y_{n+u}, Y_n^{(k)}, X_n^{(l)}) \log \frac{p(Y_{n+u} | Y_n^{(k)}, X_n^{(l)})}{p(Y_{n+u} | Y_n^{(k)})} \quad (9)$$

Where u is the prediction time, usually u=1,1=k=1. The formula (9) expresses the dependence of Y on X, these parameters can directly affect TE values.

If k and l are very large, $p(Y_{n+1}, Y_n^{(k)}, X_n^{(l)})$ will become sparse and dispersed, so even there is no causal relationship, transfer entropy will still increase automatically. Therefore, Gourévitch et al. eliminated this influence by deviation and normalization [39].

For discrete time series, TE is calculated based on binning method [32, 34]. For continuous time series, different techniques have been proposed to estimate the TE, e.g Gaussian model [45-46], kernel function based on correlation integral, and Kraskov, Stogbauer, and Grassberger (KSG) algorithm [45]. TE based on Gaussian model should meet the joint Gaussian distribution to analyze linear interactions among multivariate stochastic processes. Both Kernel function and KSG are suitable for linear and nonlinear time series and model-free, and KSG is non-parameter estimation which accuracy is higher than kernel method.

Transfer entropy is an effective metric for linear or nonlinear causal relationship between two random variables [47-48]. But there may have a common input, or existing intermediate variables. Therefore, multivariate transfer entropy is essentially based on conditional TE. Assume that the source X, objective variable Y and condition variable Z, where Z is the set of multi-variables, $X_n^- = [X_n X_{n-1} X_{n-2}...]$, $Y_n^- = [Y_n Y_{n-1} Y_{n-2}...]$, $Z_n^- = [Z_n Z_{n-1} Z_{n-2}...]$ represent the past states of X, Y and Z. Then, the equation of multivariate transfer entropy from X to Y conditioned on Z is expressed as:

$$TE_{X \to Y|Z}(k,l) = \sum p(Y_{n+1}, Y_n^{(k)}, X_n^{(l)}, Z_n^-) \log \frac{p(Y_{n+1} | Y_n^{(k)}, X_n^{(l)}, Z_n^-)}{p(Y_{n+1} | Y_n^{(k)}, Z_n^-)} \quad (10)$$

When the number of variables is more than three, Z can be extended to the representation of multivariate.

Multivariate TE could lead to "curse of dimensionality" which may produce a higher computational load, and reduce the reliability of the results [48-49]. Jakob [49] proposed the concept of transfer entropy decomposition, which based on PC-algorithm improved to calculate multi-variable transfer entropy. Lizier [50] proposed the multi-variables transfer entropy based on greedy search algorithm which was employed in this paper.

In this paper, we analyze the fluorescence image data and MEA data under three levels. Firstly, in the extracellular level, we calculated causality interactions between guard cells from optical recording electrical signals. Secondly, in the multicellular level, we found causal network in phloem by analyzing electrical signals extracted of fluorescence image of Helianthus annuus. L (H. annuus) stem tissue. Thirdly, we obtained multi-channel electrical signals of H. annuus leaf based on MEA method and found causal network of electrical signals transfer in leaf. We also used contrast experiment to validate whether there is causal network for electrical activity in leaf if the leaf petiole was killed.

We compared the causal network inferred by GC method and TE method for pairwise variables and multi-variable. Pairwise variables computing allow us to calculate all pairwise variables combination to acquire the causal network. Here, multivariate computing was performed to measure the causal interactions among the neighbor variables by iterating selecting the neighbor variables for each variable. Computation framework for causality network inference from optical recording and MEA using a hybrid method is illustrated in Figure 1.

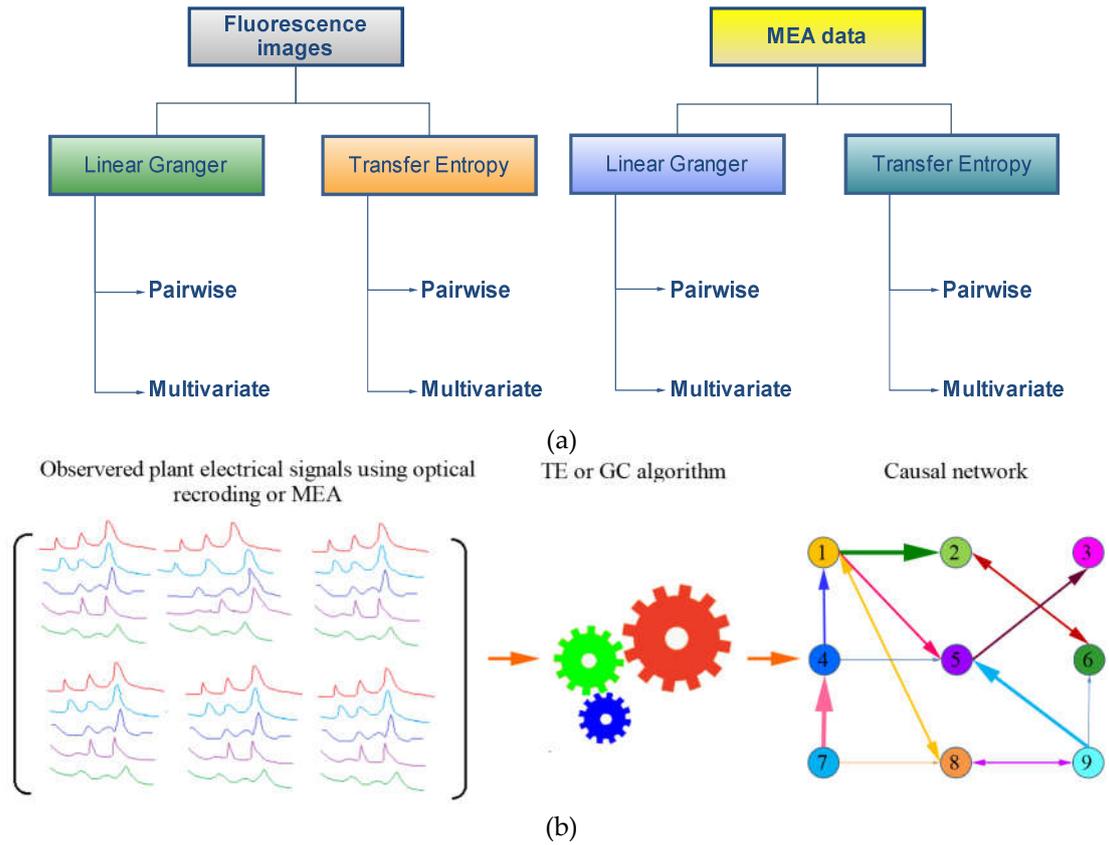

**Figure 1.** Computation framework for causality network inference from optical recording and MEA using a hybrid method (a) Computing methods for optical recording data and MEA data (b) Sketch of data flow using the proposed methods.

The hybrid causality analysis algorithm for multivariate is following. Here, we defined the adjacent regions of the selected region as the regions which are adjacent in space. As shown in the Figure 2, the adjacent regions of number 6 are {1, 2, 3, 5, 7, 9, 10, 11}. Hence, the number of adjacent regions is commonly 3, 5, and 8.

| 1  | 2  | 3  | 4  |
|----|----|----|----|
| 5  | **6**  | 7  | 8  |
| 9  | 10 | 11 | 12 |
| 13 | 14 | 15 | 16 |

**Figure 2.** An example of the adjacent regions

| **Algorithm 1. Multivariate causality matrix computing** |
|---|
| **Input: Raw electrical signals from multi-regions** |
| **Output: Causality matrix** |
| 1. The recording area were equally divided into a grid with *R* rows and *C* columns. The total number of small regions in the grid is represented by *N* which equals to *R\*C*. All regions are numbered from 1 to N. The observed signals (time series) were corresponding to the numbered region in the grid.<br>2. Stationary test by ADF test. // This step is only for Granger Causality.<br>3. Calculating the multivariate causality matrix.<br>   For i = 1: N // N is the total number of time series.<br>      Find the adjacent regions for the i-th region. Assume the number of these regions as $K_i$.<br>      Let $ID_i$ as the array of the sequence number of these regions.<br>      Calculating the Causality matrix $A_i$ for the $K_i$ observed signals by multivariate GC or TE method.<br>   ENDFOR<br>4. Define N-by-N matrix *M* and initialize it with zero.<br>5. For i = 1: N // Merge all causality matrix $A_i$ (i=1,..,N) into M<br>    For a = 1: $K_i$<br>      For b =1: $K_i$<br>        x = $ID_i$ (a); y= $ID_i$ (b); // Get the sequence number of the regions<br>        IF M(x,y) < $A_i$ (a,b)    // Select the maximum value<br>          M(x,y) = $A_i$ (a,b);<br>        ENDIF<br>      ENDFOR<br>    ENDFOR<br>   ENDFOR<br>6. END |

*2.3 Characteristics of causal networks*

To get a deeper insight of the causal networks, network density and node flow were employed to evaluate the performance of network links [51-52].

The network density can be expressed as $d = \frac{1}{N(N-1)} \sum_{i=1}^{N} \sum_{j \neq i}^{N} link(j \rightarrow i)$, where *N* is the number of all nodes in network. It is the ratio of the number of existing links in network to the all potential pairwise links. Furthermore, the node density can be defined as the ratio of the number of links with node *i* to all potential *N-1* links. High value of network density reflects the globally integrated activity. And the high node density reflects that it is a hub node.

The net out-degree of node *i* is the difference of the out-degree and in-degree. The positive net out-degree means the node *i* is the source node. The negative net out-degree means the node *i* is the sink node. Considering the causal network in the paper, we use causal density to refer the network density, and causal flow to refer the net out-degree.

## 3. Experiment results and discussion

*3.1 Datasets*

3.1.1 Optical recording datasets

To estimate the causality and information flow of the signals by GC and TE analysis, the observed data of plant electrical activity with higher spatial resolution by optical recording were based on our previous work [11], which recording methods were described in details. The electrical signals extracted from series of consecutive raw fluorescence images of three guard cells in an intact leave of *Helianthus annuus. L* (*H. annuus*) induced by heat stimulation.

492 fluorescence images were acquired by a common florescence microscope with a 40×objective lens at 5fps sampling rate. The electrical signals induced by heat stimulation were transferred from the petiole or the lower part to leaves. We divided every images into 13 columns*10 rows, totally 130 rectangular regions. There are 130 rectangular regions in image, as shown in figure 1(a). We calculated GC and transfer entropy of electrical signals in guard cell to exploit dynamic information transfer among the cells and within the single cell.

In the second application case, data were from our published paper (the raw dataset for Figure 2 in that paper) [11].The total measurement area under the microscope was 0.435 × 0.328 mm$^2$ in stem using a 20 × NA = 0.40 objective lens. 128 fluorescence images corresponding to electrical activity induced by electrical stimulation were obtained at 1fps sampling rate. Each image was divided into 20*27=540 regions, and then 540 raw signals (time series) were extracted from 128 fluorescence images shown in Figure 1(b). We calculated GC and transfer entropy of action potentials signals in *H. annuus* stem to exploit dynamic information transfer in plant stem.

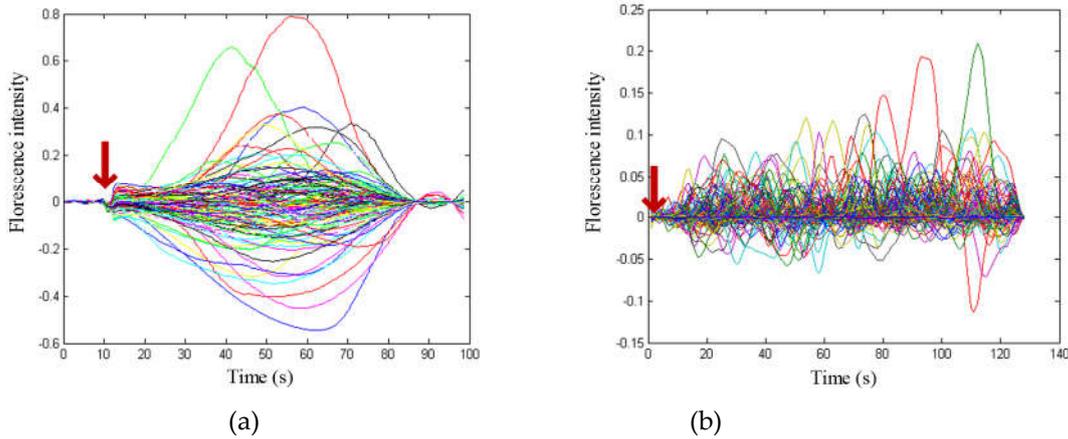

(a)  (b)

**Figure 3.** (a) 130 electrical signals of different regions of fluorescence images of guard cells, stimulation time is at 9.6s. Thermal stimulation is performed at 9.6s and the end of artifacts of thermal stimulation is at 13.6s. (b) 540 electrical signals extracted from fluorescence images of sunflower stem. Stimulation time pointed by red arrow. The arrow indicates the timing of the stimulation in (a) and (b).

3.1.2 MEA datasets

In the third application case, the raw MEA datasets are from our previous work [14].The MEAD64 array recorded the heat-induced electrical signals in *H. annuus* leaf by 64 electrodes. The first MEA dataset is corresponding to Figure S2 in supplemental information in our published paper [14]. The dataset consists of spatio-temporal propagation of electrical signals in a leaf. Also, the second MEA data is corresponding to Figure S4 in supplemental information in that paper. The dataset is made up of recordings induced by heat in leaf after leaf petiole was killed by 10% paraformaldehyde.

The two different datasets were analyzed and compared by GC and TE methods. The analyzed results will be used to verify the consistence between the electrical signals dynamic causal network and the raw wave propagation, and to visualize the dynamic causal network in leaf.

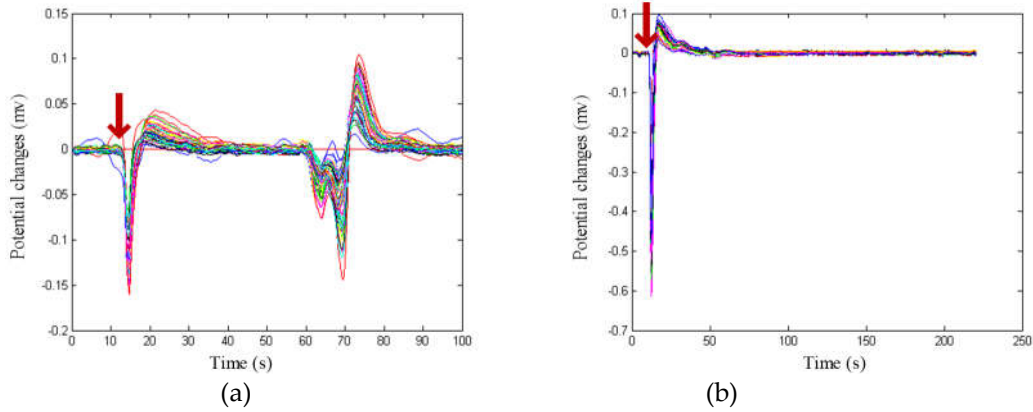

**Figure 4.** (a) Variation potentials of MEA recording by heat stimulation, the heat stimulation occurred at 12s. (b) MEA data after the leaf petiole was killed by 10% paraformaldehyde, the heat stimulation occurred at 10s

*3.2 Data preprocessing and parameters selections*

In the GC analysis, the stationarity and linearity in plant electrical signals should be identified at first. The Augmented Dickey Fuller (ADF) test was used to test stationarity. The variation potential of guard cell was covariance stationary by ADF test. However, the AP signals by optical recording in the phloem region are non-stationary identified by ADF test, then second -order differencing were applied to AP signals to meet stationary. For the first MEA dataset, it was not stationary, and second-order differencing was also carried out to meet stationary. The second MEA dataset is also non-stationary, but first-order differencing is enough to meet stationary.

Nonlinearity test can be used to determine linear GC model or nonlinear GC model. We adopted the nonlinear test referring to the literatures [44]. The results showed that all data were not nonlinear. Thus, we selected the linear GC model. To determine the model order, both the Bayesian information criterion (BIC) and Akaike information criterion (AIC) were used. The minimal value of them was chose as the order of model. For GC method, the chi-square test was used, and $p < 0.01$ was considered significant.

In transfer entropy method, the plant electrical signals data are continuous and they are not in accord with Gaussian-distribution model. So we selected the KSG kernel in transfer entropy. The number of nearest neighbor points is 4. For TE method, the empirical bootstrap test was used, and $p < 0.01$ was considered significant. The delay is 1 point because we just focused on the causal interactions among adjacent regions.

*3.3 Results and analysis*

3.3.1 Causal network inference for electrical signal of guard cells

(1) Pairwise GC causal network inference

We computed the pairwise causal interactions among the 130 regions for three guard cells by both GC and TE methods. Figure 5 shows the dynamic causal networks at 20 seconds (baseline), 48 seconds (in the depolarization process after stimulation), and 88 seconds (in the repolarization after stimulation). The first row is the pseudo-color maps of GC matrixes. And the GC magnitude decreased over time. However, the map of GC network connections in second row increase over time. This phenomenon is according with the depolarization and repolarization process.

The increasing causal network connections coincide with the gradual enhancement of plant electrical signals transfer network in the regions of guard cells. In the connection networks, the bi-direction and unidirection connection both exist. Some areas are neighbors but without causality connections.

The dynamic network connections show the formation process of causal network. Firstly, the connections appear not only at each individual guard cell membrane itself but also at the neighbor regions of 3 guard cells. Next, each guard cell has connections in the cell membrane and they also connected together. Then, the connections extend to other regions.

The pseudo-color maps of dynamic causal density (the first row) and causal flow (the second row) are shown in Supplementary Figure 1. Based on the causal density per node, we can infer which node has more connections with its neighbors. From the node causal flow, we can find which the node is the source or sink node.

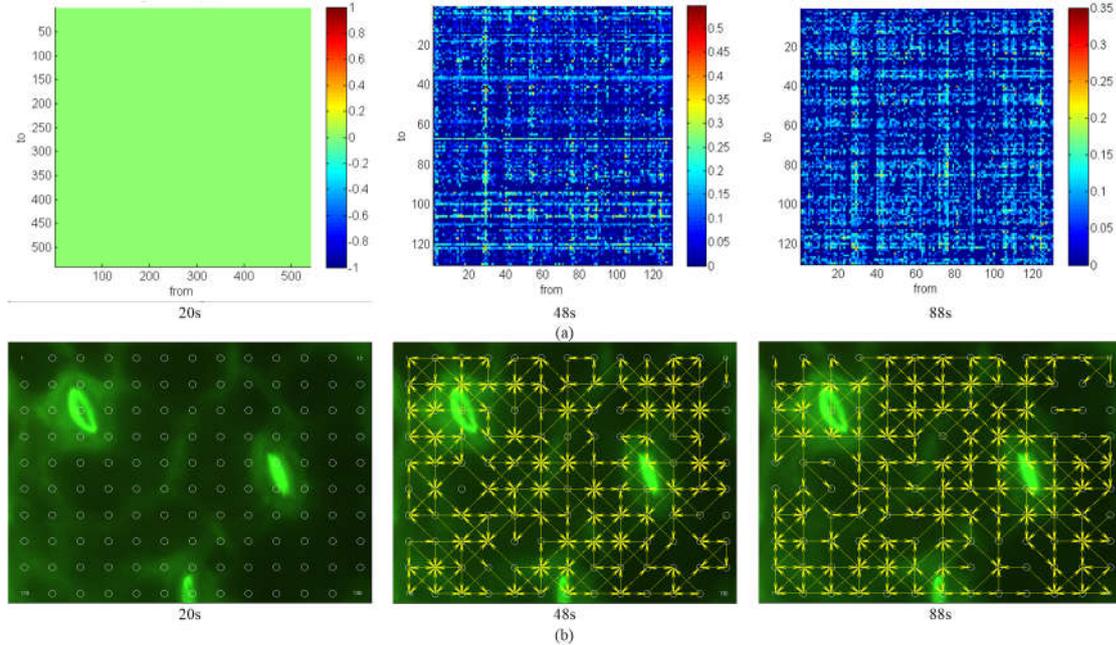

**Figure 5.** Pairwise Granger causality network of guard cells（a）Pseudo-color maps of the causal matrix at individual snapshots using GC method. (b) Dynamic causal networks at individual snapshots in time. 20s：baseline，44s, in the depolarization process after stimulation, 88s: in the repolarization after stimulation.

Figure 6 shows the dynamic causality networks inferred by TE. Similar to GC methods, the network connections became increasingly intensive over time at 20s, 48s, and 88s. And the TE matrix values decrease slowly as time increase. Compared to GC, the network connections at 8 seconds in TE are more intensive than GC network connections. However, there were no connections at 20 seconds at Figure 5. The results show that pairwise TE method is more sensitive to small perturbations in data. The pseudo-color maps of dynamic causal density (the first row) and causal flow (the second row) are shown in Supplementary Figure 2.

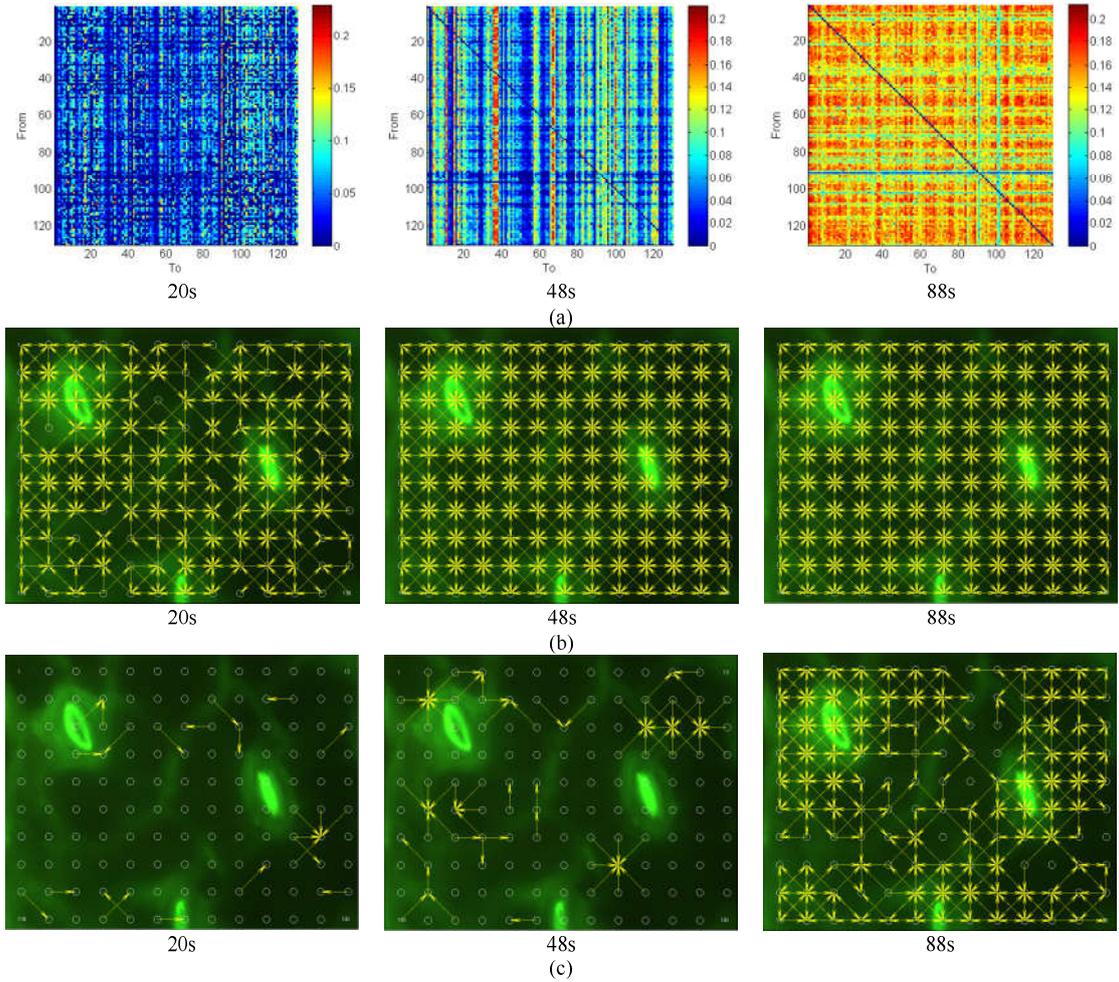

**Figure 6**. Pairwise TE network of guard cells (a) Pseudo-color maps of the causal matrix at individual snapshots in time using TE method. (b) Dynamic causal networks at individual snapshots in time with a threshold = 0. (c) Dynamic causal networks at individual snapshots in time with threshold = 0.14. 20s：baseline，44s, in the depolarization process after stimulation, 88s: in the repolarization after stimulation.

(2) Adjacent multivariate causal network inference

There may be spurious links between pairwise causality interactions. Therefore, we consider multivariate causal network inference in adjacent regions. The results are shown in the Supplementary Figure 3 and 5. The results based on multivariate GC network inference are also according with the depolarization and repolarization process. Multivariate TE network is successfully reducing the false links. The dynamic causal networks show that the connections are simultaneously and mostly appearing in the middle regions of the three guard cells at 22s. Then all regions began to generate connections at 48s. However, it may also be affected by the noise data for the existing many redundancy links. The characteristic estimation of causal networks for both GC and TE are shown in Supplementary Figure 4 and 6.

3.3.2 Causal network inference of APs using optical recording in *H. annuus* stem

(1) Pairwise causal network

As shown in Figure 7, pairwise GC interactions networks at time 40s, 80s, and 128s were indicated respectively, including the pseudo-color maps of pairwise GC matrixes (the first row in Fig.6) and the corresponding connection networks (the second row). The connections increased gradually in both the pseudo-color maps and the connection images over time, but the value of GC influence decreased rapidly in the repolarization process of electrical signals. At 40s, the GC matrix value is 0, indicating that there is no network connections. The results show that, the network connections are increasing gradually over time in the middle of phloem in stem. Finally, the connections are not only appearing in the longitudinal direction of phloem, but also in the right cross section of phloem. The findings strongly support the inferences about the electrical activity pathway not only in longitudinal direction but also in the lateral direction [53].

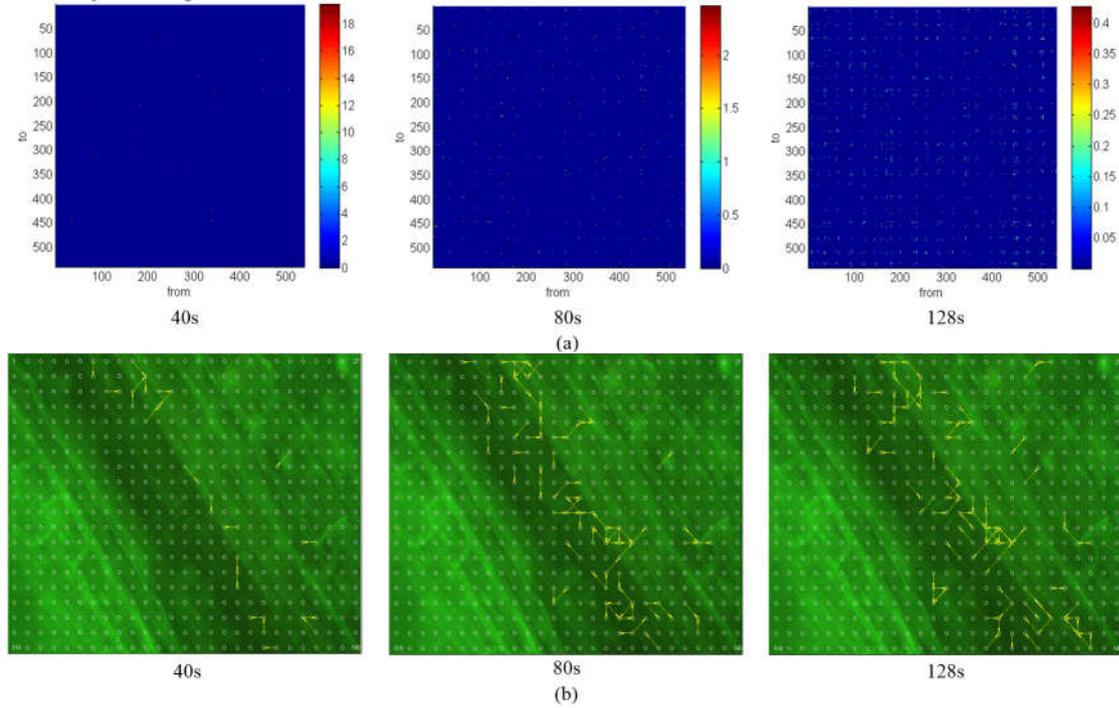

**Figure 7.** Pairwise GC causal network of *H.annuus* stem (a) Pseudo-color map of GC matrix at individual snapshots. (b) Dynamic causal networks at individual snapshots.

The results of TE for pairwise varieties at time 40s, 80s, and 128s are shown in Fig.8, which includes the pseudo-color maps (the first row), the corresponding connection images with a threshold at 0 (the second row). The results of TE also show that network connection increases gradually with the increase of time. Compared with the GC, TE is more sensitive to noise data, and the density of network connection of TE is larger than GC. The characteristic of pairwise causal networks for both GC and TE are shown in Supplementary Figure 7 and 8.

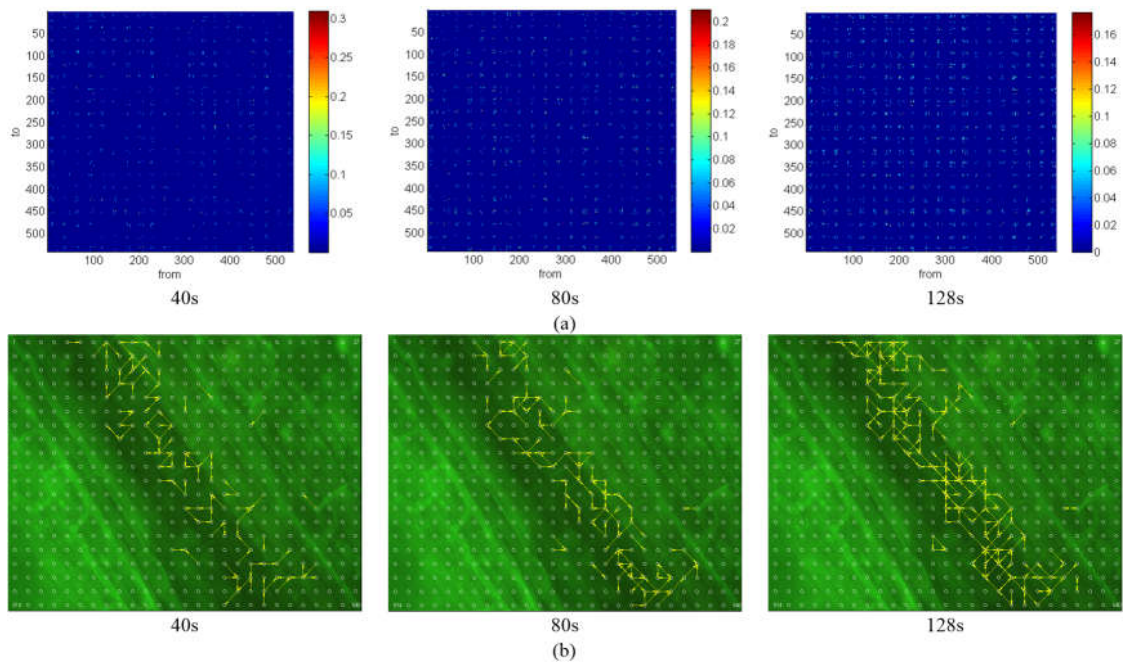

**Figure 8**. Dynamic pairwise TE causal network of paired variables of *H.annuus* stem (a) Pseudo-color maps of TE causal matrixs (b) Dyanmic TE causal networks at individual snapshots in time with a threshold =0.

(2) Adjacent multivariate causal network inference

To avoid the spurious links between pairwise causality interactions, we applied multivariate causal network inference in adjacent regions. The results are shown in Supplementary Figure 9 and 11. The dynamic multivariate GC causal networks are sparse for the decrease of spurious links. However, the multivariate TE causal networks are still intensive. The results of multivariate GC causal networks are also consistent with the plant signals transfer in the phloem though with sparse links. But the TE causal networks are still full of redundancy links. Also, characteristics of the causal network are shown in Supplementary Figure 10 and 12.

3.3.3 Causal network inference of electrical signals in *H.annuus* leaf by thermal stimulation
(1) Pairwise causal network

Figure 9 shows the corresponding GC results at time 55s, 75s and 100s after the thermal stimulation. With the increase of time, the network connection gradually increases. Finally, the magnitude values of GC, i.e. the connection strength decreases. Because the region area of MEA recording is greater than the above optical recording, which can indicate the information transfer and the dynamic formation process of causal networks among multiple cells or tissues. The network estimation are shown in Supplementary Figure 13.

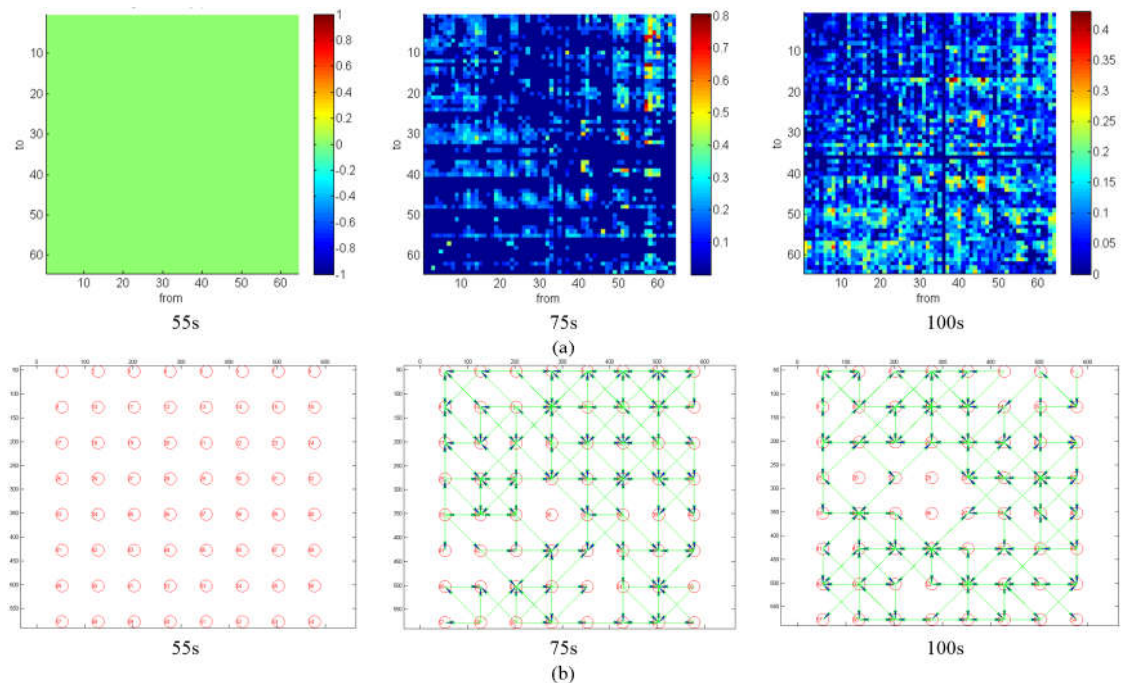

**Figure 9.** Dynamic pairwise GC network of *H.annuus* leaf after burning stimulus (a) The pseudo color maps of causal matrix at different time. (b) Dynamic GC causal networks at different time.

Figure 10 shows the corresponding transfer entropy results at 55s, 75s and 100s after the thermal stimulation. The TE results are similar to the GC results. The transfer of information within the leaf regions without the generation of electrical signals was also calculated by TE. The result showed that the transfer entropy is more sensitive to subtle fluctuation of baseline. While the Granger method can better reflect the connection state of network before electrical signals are induced. The characteristics of network are shown in Supplementary Figure 14.

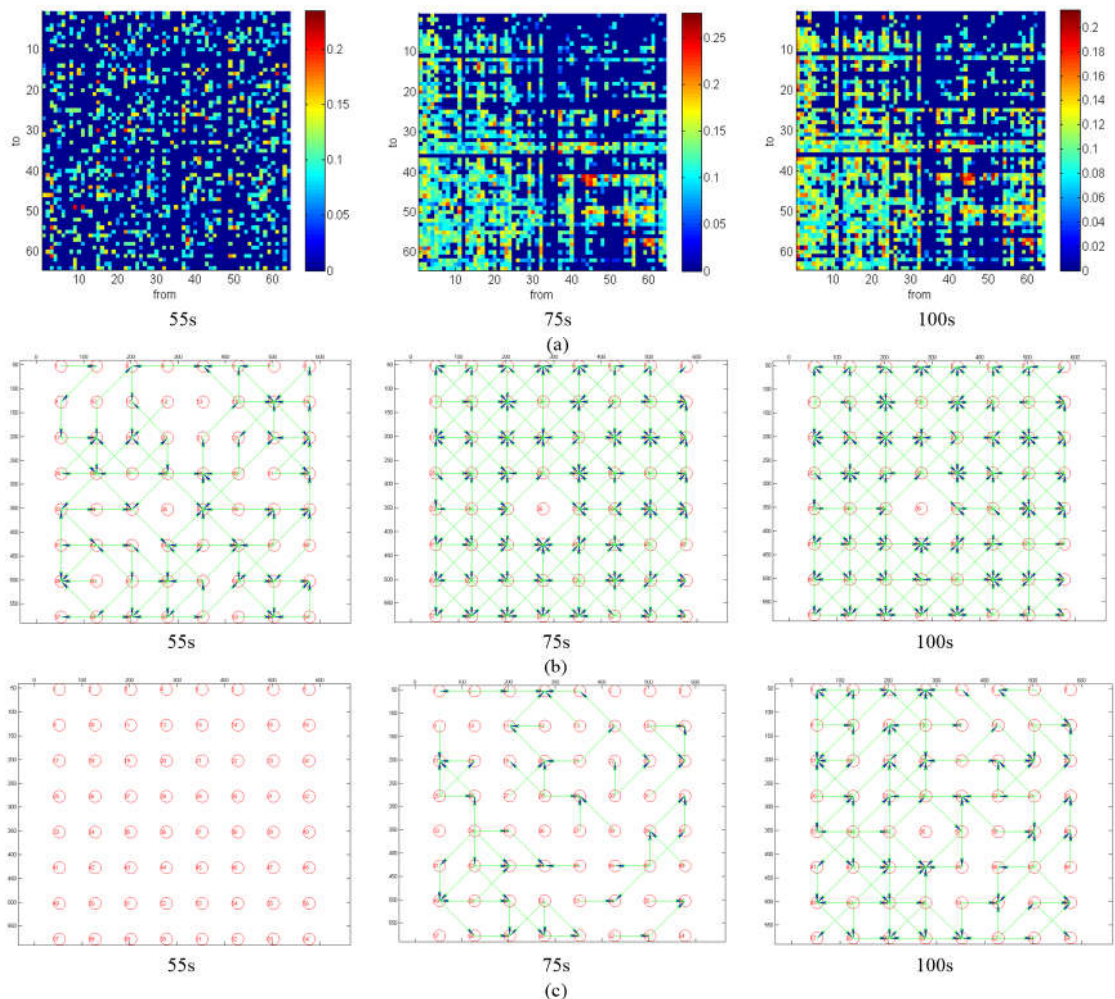

**Figure 10.** Dynamic pairwise TE causal network of *H.annuus* leaf after burning stimulus (a) Pseudo color maps of TE causality matrix at different time. (b)The transfer entropy causal networks at different time with a threshold =0. (c) Dynamic TE causal networks at individual snapshots in time with different threshold at 0.2, 0.15, and 0.1.

(2) Adjacent multivariate causal network inference

The multivariate causal network are shown in Supplementary Figure 15-18 by both GC and TE methods. For multivariate GC causal networks, the results show similar results with the pairwise causal network. But multivariate GC causal networks are very sparse. Multivariate TE causal networks are very intensive which is similar to pairwise TE causal networks. The network estimation are also shown in Supplementary Figure 16 and 18.

3.3.4 Causal network inference of MEA data using 10% paraformaldehyde
(1) Pairwise causal network

We computed the pairwise causal interactions among the 64 channels MEA data by both GC and TE. Figure 11 shows the dynamic causal networks of GC matrixes at different time at 125s, 162.5s, and 200s. The first row shows the pseudo-color maps of GC matrixes, and the second row are the causal networks of GC over time. After the leaf petiole was killed by 10% paraformaldehyde, the variation potential signals cannot transfer to the leaf which lead to no VP appeared in the leaf. There are few connections among the three causal networks. Hence, it was consistent with the biological explanation. The characteristics of GC causal networks are shown in Supplementary Figure 19.

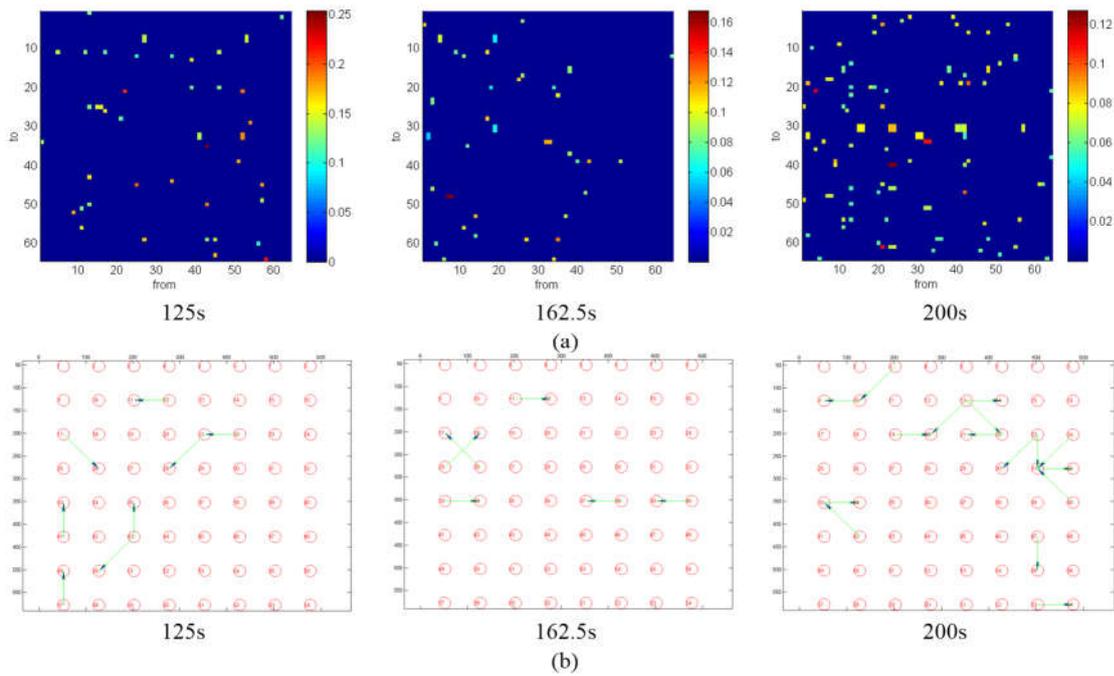

**Figure 11.** Pairwise GC networks of *H.annuus* leaf after the leaf petiole was killed (a) Pseudo-color maps of the GC matrixes at individual snapshots in time. (b) Dynamic causal networks at individual snapshots in time.

The results of TE are shown in Figure 12 at time 125s, 162.5s, and 200s. The pseudo-color maps and causal networks are more intensive than to GC results. But the connections density is much smaller than the results in Figure 12. Therefore, the results of TE are also satifying the biological explanation. The characteristics of TE causal networks are shown in Supplementary Figure 20.

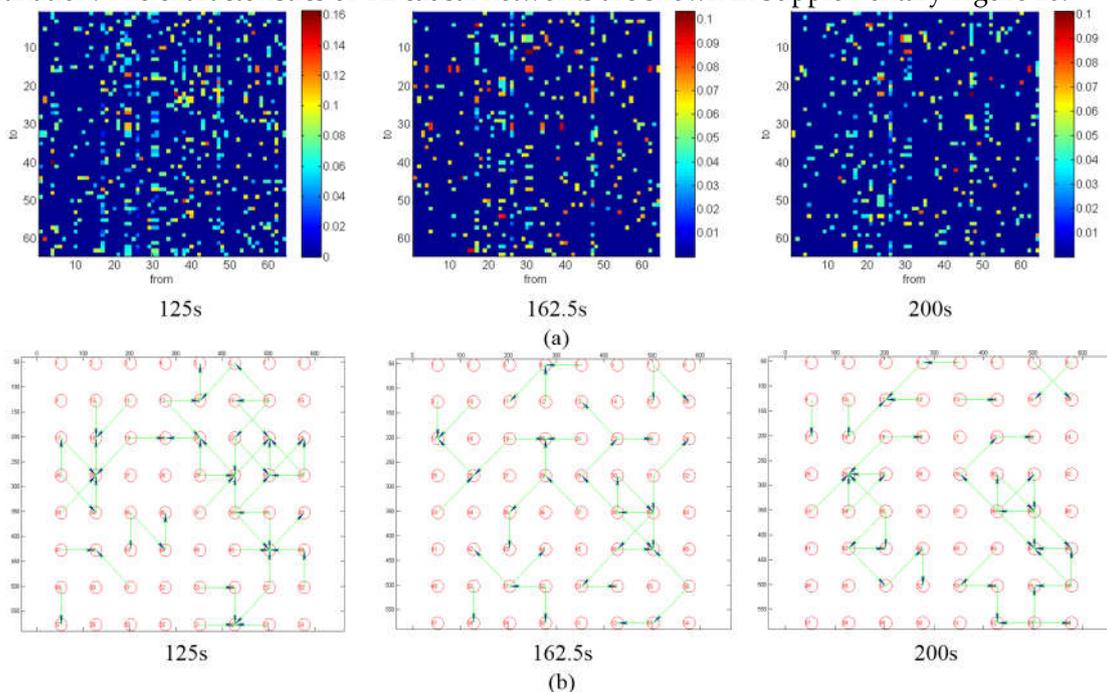

**Figure 12.** Pairwise TE networks of *H.annuus* leaf after the leaf petiole was killed (a) Pseudo-color maps of the TE matrixes at individual snapshots in time. (b) Dynamic causal networks at individual snapshots in time.

(2) Adjacent multivariate causal network inference

The results of multivariate GC causal network are shown in the Supplementary Figure 21. Multivariate GC results are consistent with the biological explanation. However, the multivariate TE networks were not calculated because the time series are in baseline state without significant change.

And the tiny variation in the signal amplitude is induced by noise in fact. The characteristics of GC causal network are shown in Supplementary Figure 22.

*3.4 Discussion*

TE method is more sensitive to noise when the signals are relatively steady and the variations are induced by noise, which may lead to wrong results [54-56]. Especially during a resting state, the result using pairwise TE indicated spurious connections with a threshold=0 (see Fig.6). Despite existing many causal links, but most of them are weak links in rest state of plant, namely this result is affected by the noise in resting state of plant. To obtain a real dynamic TE effective networks, we set a proper threshold at 0.14 for TE values. Then the dynamic TE effective networks in Figure 6(b) shows consistence with the plant electrical signals transfer. In depolarization and repolarization, the number of links will increase and become stronger. This phenomenon shows dynamic links among plant cells. When in reset state, there only existing few links, but after the stimulation, the number of links and their magnitude will increase, then the information flow into all regions. In addition, the TE value gradually decreased during the process, it implied that the threshold was time dependent (see Figure 6, Figure 10, and Supplementary Figure 16). There are few reports for setting threshold using TE analysis [57-60]. However, how to the set an adaptive threshold using TE analysis is worth studying in the future.

Compared to the TE analysis, GC method is not sensitive to noise in our cases. During resting state, though recordings are mainly induced by noise data, the corresponding GC networks links are few or even none. When stimulation is used, the links become increasing and stronger. In depolarization process, the plant electrical signals have relatively large changes which also lead to large GC values. In repolarization process, the plant electrical signals start to recover to the resting state with gradual decreasing variation, which lead to a decreased GC values. In the process, the number of links is still steady.

Both TE and GC can be used to infer the effective network of plant electrical signals. GC is better to reflect the dynamic changes of effective networks, but the data should be stationary. TE method can also applied into discovery of effective networks. However, it is sensitive to noise data. Threshold is essential to exclude the false links to discover the important links in effective networks. TE and GC can two supplemental methods in effective networks inference.

Bidirectional links in effective networks imply that the coupled causal interactions among some regions. The unidirectional links are also important properties in effective networks which show the special direction of information flow. In addition, our result demonstrated that the changes in direction of some links in the dynamic effective networks during the depolarization stage are different from in repolarization stage.

For pairwise TE or GC method, the results can reflect the local links changes of every region in meticulous way. However, there are a few spurious links among the pairwise connections especially for signals with noise. The multivariate GC method is based on the conditional variables which will help to decrease the false links or indirection links. Consequently, the number of links in multivariate effective networks is less than the pairwise effective networks in GC methods. The results shown in Supplementary Figure 3, 9, 15, 21 can also reflect the transfer pathway of plant electrical signals. In our experiences, it took more than one hundred hours to measure the causality using multivariate TE method running in a single PC. Consequently, a high performance computing cluster is recommended by Joseph Lizier.

The four datasets together revealed the generation of effective networks in three levels in plant cells including the single cell level, the multicellular level, and the tissues level. In cell level, it allows us to see membrane potential wherever response occurs and at the sites in which spikes are initiated, and know how a plant cell integrates its response input into its electrical signal output. Both TE and GC discovered the formation and transfer process of plant electrical signals around signal cell. Next, in multicellular level, the dynamic effective networks show the plant electrical signals transfer in the phloem cells region. In the tissues level, combined MEA datasets showed the formation and degradation process of effective networks in the plant leaf. In other words, our methods have provided more kinds of information flows about the propagation of the plant electrical signals that were unobtainable.

In fact, there is distinct evidence that plants have several closely interconnected signaling pathways in plant biology. However, at present we still lack information on the integrated timing of hydraulic, electrical, and chemical signaling pathways, and the importance of each pathway for information transmission. In the future, our proposed computing method is a potential powerful tool for the analysis of complex effective connectivity of these signaling pathways.

Beyond this topic, we believe that Granger causality combined with transfer entropy computing method can extend and develop these frameworks for novel problem domains in other fields, e.g. social network, business network.

## 4. Conclusion

In this paper, both GC and TE can provide overall descriptions of causal networks of plant electrical signals. Pairwise and multivariate causal networks were inferred based GC and TE which showed similar dynamic changes of network links over time. Furthermore, our analysis based on GC and TE for three qualitatively different levels of plant bioelectrical activities revealed direction of information flow and dynamic complex causal connections to understand the relationship between bioelectrical activity and its physiological response. Besides the visualization of causal networks, the visualization of node density and the net out-degree were important measures to discover the hub nodes, source nodes and sink nodes in the causal flow of causal networks. Although there are limits in applying the GC and TE methods, but they are still useful tools if the parameters are set carefully. Our work provides an initial insight of the plant electrical signals spatio-temporal networks. Future work will include function networks inference, multivariate TE methods with proper threshold setting, reducing the spurious links, using binning transfer entropy, and reducing the influence of noise data on the transfer entropy, etc.

**Acknowledgments:** We thank for thank Dr. Joseph T. Lizier for the discussion on the transfer entropy analysis. This research was supported by the National Natural Science Foundation of China (61571443), the Specialized Research Fund for the Doctoral Program of Higher Education (20130008110035), and the National Key Scientific Instrument and Equipment Development Projects (2011YQ080052). The authors would like to thank the Key Laboratory of Agricultural Information Acquisition Technology of the Chinese Ministry of Agriculture for their support.

**Author Contributions:** Yang Chen and Lan Huang analyzed the data and design the algorithms. Dong-Jie Zhao and Zi-Yang Wang conceived, designed the experiments and performed the experiments; Yang Chen, Zhong-Yi Wang and Lan Huang wrote the paper. Zhong-Yi Wang, Chao Song, and Wei-He Liu and Guiliang Tang revised the manuscript. All the authors read and approved the final manuscript.

**Conflicts of Interest:** The authors declare no conflict of interest.